**Is the annualized compounded return of Medallion over 35%?**

Shuxin GUO
*School of Economics and Management*
*Southwest Jiaotong University*
*Chengdu, Sichuan, P. R. China*
shuxinguo@home.swjtu.edu.cn
ORCID: 0000-0001-8188-6517

Qiang LIU
*Institute of Chinese Financial Studies*
*Southwestern University of Finance and Economics*
*Chengdu, Sichuan, P. R. China.*
qiangliu@swufe.edu.cn
ORCID: 0000-0001-8466-3108

**Acknowledgements**

This work was supported by the National Natural Science Foundation of China under Grant numbers 71701171 and 72371208.

# Is the annualized compounded return of Medallion over 35%?

**Abstract**: It is a challenge to estimate fund performance by compounded returns. Arguably, it is incorrect to use yearly returns directly for compounding, with a reported annualized return of above 60% for Medallion for the 31 years up to 2018. We propose an estimation based on fund sizes and trading profits and obtain a compounded return of 31.8% before fees. Alternatively, we suggest using the manager's wealth as a proxy and arriving at a compounded growth rate of 25.6% for Simons for the 33 years up to 2020. We conclude that the annualized compounded return of Medallion before fees is probably under 35%. Our findings have implications for correctly estimating fund performance.

In *The Man Who Solved the Market: How Jim Simons Launched the Quant Revolution*, the annualized compounded returns of Simons' Medallion fund are reported to be as high as 66.1% (before fees) and 39.1% (after fees) for the 31 years from 1988 to 2018, respectively. "[A] compound return of 63.3%" is later given in Cornell (2020). Are such huge returns reliable? Very unlikely. Medallion started with $20 million in 1988 (Zuckerman, 2019). With an annual 66.1% compounding, Medallion would have grown to $374 trillion by the end of 2020, a mind-boggling amount![1]

Even though annual fund net returns[2] can be found trivially (from letters to investors), the annualized compounded return cannot be estimated easily. The reason is simple but subtle: the returns have not been fully accumulated or, in financial terms, correctly compounded.

It is a challenge to estimate fund performance by compounded returns. One important reason is that fund sizes are usually discontinuous due to cash inflows and/or outflows. As Table 1 later shows for the Medallion fund, there were multiple jumps in fund sizes from 1988 to 2001. Further, the size was mostly around 5.2 billion between 2002 and 2009, while capped at 10 billion from 2010 to 2018. How can we compute the correct compounded returns with discontinuous sizes? This is an exciting question for us to explore.

In this short note, we first show why compounding using yearly net returns can be wrong for the discontinuous value process. Then, we propose an approach based on the initial fund size or value, total trading profit, total cash inflow, and the final fund size for estimating a compounded return for a fund. Further, we suggest that the growth rate of a manager's net wealth can be utilized as a proxy for the compounded return of a fund.

**Naïve Compounding**

[1] The US equity market is $51 trillion at the end of 2020 (see https://siblisresearch.com/data/us-stock-market-value/).

[2] Net return (Cochrane, 2005), the difference between the ratio of final price to initial price and one, is the most commonly reported number in finance.

Because the annual net returns of a fund are usually known (i.e., reported to investors), it is tempting to compound by joining those returns together naively. Let's denote the yearly net returns by $\mu_t$. One might obtain the (naïve) average compounded return as follows:

$$\mu_{nc} = \left(\prod_{t=1}^{n}(1+\mu_t)\right)^{1/n} - 1 \qquad (1)$$

Each year, both the trading profit and fund size are usually reported by fund managers. Consequently, a fund's net return is simply the ratio of trading profit to the fund size at the beginning of the year. For Medallion, we show fund sizes and trading profits between 1988 and 2018, taken from Appendix 1 of Zuckerman (2019), in Table 1. Using Formula (1) and net returns in Table 1, we obtain a naïve compounded return of 63.2%, quite close to the 63.3% of Cornell (2020). The average return reported in Zuckerman (2019) is somewhat higher, at 66.1%, which is probably the simple average of yearly net returns.[3] Zuckerman (2019) uses "returns before fees," which are slightly different from the net returns in Table 1 for some of the early years within the sample period.

Table 1 here

With a compounded return of 63.2%, 20 million, the fund size of Medallion at the beginning of 1988 (Table 1) would become a staggering 78 trillion at the end of 2018. This is doubtlessly an absurd amount of money and wrong egregiously because the total trading profit of Medallion for 31 years up to 2018 is only a little bit over 100 billion (see Table 2 later).

Unfortunately, we cannot combine the yearly net returns in this way or use Formula (1) to obtain a correct annualized compounded return because the fund size of Medallion is discontinuous. We use a simple numerical example to illustrate this point.

[3] We obtain an average annual return of 66.0%, using the formula $\bar{\mu} = \frac{1}{31}\sum_{t=1}^{31}\mu_t$ with returns in Table 1.

Assume the size is capped at 100 at the beginning of each year, with a 50% net return for two years in a row. The initial 100, with two 50 profits, will grow to 200.[4] In this case, the final value of 200 is the sum of yearly profits and the initial capital. The (annualized) compounded return is computed as $(200/100)^{1/2} - 1 = 41.4\%$, which is significantly lower than the naïve compounded return of $[(1 + 50\%)(1 + 50\%)]^{1/2} - 1 = 50\%$.

If compounded at 50%, 100 will grow to 225 in two years, which is incorrect. With a 50% return for ten years in a row, the annualized compounded return would be much lower at 19.6%, while the naïve compounded return would still be 50%. Clearly, the longer the investment horizon, the bigger the error of the naïve compounded return at measuring the actual compounding effect. For 17 years up to 2018, the sizes of Medallion had been fixed at two different levels (see Table 1). Therefore, the naïve compounded return will be incorrect as a measure of Medallion's performance.

**Compounding on Profit Reinvested**

Compounding means earning interest on interest. It is common sense that to compound, we must keep the profit earning more profit. Let's look at a numerical example.

Assume Year 1 grows from 100 to 150, and Year 2 does from 200 to 300. Note that at the beginning of Year 2, there is an outside investment (i.e., cash inflow) of 50. Out of the profit of 100, the share of the 150 capital is 75. Consequently, the 100 at the beginning of Year 1 grows to 225 at the end of Year 2, and the compounded return is $(225/100)^{1/2} - 1 = 50\%$. The naïve compounded return is also 50% and correct!

Therefore, when profits are reinvested, the naïve compounded return can be correct, even if there are extra outside investments (or discontinuous fund increments).

**Size-Profit Compounding**

---

[4] The profit of the first year that is not reinvested can earn interests in principle. It can be complicated to keep track of money taken out of the fund i compounding the initial investment. Therefore, interest is ignored in this short paper.

From the previous analysis, we propose to estimate an annualized compounded return for an *N*-year fund as follows without considering the time value of money:

$$\mu_c = \{\frac{1}{V_1}\left(V_1 + P_1 + \sum_{t=2}^{N} P_t \min\left[1, \frac{V_{t-1} + P_{t-1}}{V_t}\right]\right)\}^{\frac{1}{N}} - 1 \qquad (2)$$

where $V_t$ is the initial capital (i.e., fund size) at the beginning of year *t* and $P_t$ denotes the year-*t* trading profit. If there are extra outside investments after the first year, profits must be shared with additional investors, and therefore, the fund (for the purpose of compounding) is allocated the part of profit that is proportional to the prior-year sum of size and profit. This treatment may overestimate or underestimate the fund's total profit. For example, the fund size jumped up to 42 in 1991 and then up to 74 in 1992 (Table 1), but the profit in 1992 is adjusted on the sum of size and profit in 1991, both of which contain outside investments. As a result, the allocated profit is overestimated. On the other hand, profits taken out of the fund without reinvestment can earn interest and lead to a higher return than Formula (2).

Using Formula (2), we can adjust the trading profits and compose a hypothetical compounded year-end value series for the initial capital in 1998 (Table 2). It is clear that the (compounded) gross returns (Cochrane, 2005) decreased around 2010, when the fund size was large and capped at 10 billion with very high annual net returns. Utilizing Formula (1) on gross returns or Formula (2), we obtain an annualized compounded return of $31.8\%$. The corresponding log return (Cochrane, 2005) is $27.60\%$, with an annual volatility of $19.65\%$. Consequently, the modified "Sharpe ratio" as suggested by Guo and Liu (2024) is $27.60/19.65 = 1.40$.

Table 3 shows estimated annualized compounded returns over five sub-periods for the Medallion fund. Three observations can be made. First, the return becomes lower with bigger fund sizes. The return for the first 14 years is high at $51.8\%$, but for the last 17 years relatively low at $19.1\%$. Similarly, the return between 2002 and 2009 is higher than that between 2010 and 2018. Second, the return is lower for longer periods for similar sizes. The returns between 2002 and 2009 and 2010 and 2018 are higher than that between 2002 and 2018. Third, higher returns are due to the initial tiny size of the fund. The returns over the first 14 and the whole 31 years are higher than the other three sub-periods.

Table 3 here

Further, the Medallion fund may have a maximum capacity of 10 billion because it has been capped at that level for almost a decade as of 2018. Therefore, it can be predicted that the annualized compounded return for the 2010-2018 sub-period will go down further in the long run if the yearly profits in the future are going to be at the same level on average.

Our estimated annualized compounded return of Medallion for the 31 years, 31.8% before fees, is undoubtedly awe-inspiring and remarkable. It is less than 35% however, and significantly lower than the naïve compounded return of 63.2%, or those reported previously by others (Zuckerman, 2019; Cornell, 2020).

**Net Wealth Compounding**

Another possible approach is to use the growth of Simons' net wealth as a proxy because many media companies track the total net assets of wealthy people. According to Forbes, Simons had a net worth of 23.5 billion as of November 16, 2020, after charitable givens of 2.7 billion. In other words, Simons had a total asset of 26.2 billion.

Simons' initial net wealth is not easy to know, unfortunately, but a rough estimation is still possible. Zuckerman (2019) writes that Simons had over 50% equity in 1998, after giving 10% to an associate (and later sizable to three other employees). Let's assume 50%. This amounted to 550 million for Simons in Medallion, which had a size of 1.1 billion at the beginning of 1998 (Table 1). Consequently, the wealth of Simons grows by an annualized compounded rate of 18.3% for 23 years up to 2020.

Medallion started with a size of 20 million in 1988. Using the information from the previous paragraph, we assume that Simons had 70% equity (50%+10%+10%) or 14 million capital. In this case, he would have an annualized compounded growth of 25.6% for 33 years up to 2020.

Both numbers are after fees and taxes and are truly impressive! But they can be incorrect, because the initial net assets are rough estimates. Still, it is probably safe to guess that they are less than 35%! Again, the initial tiny size of wealth leads to a higher growth rate over a longer 33-year period.

It is unclear how taxes and fees together will alter Simons' returns. Even though taxes lower Simon's wealth growth, fees will inflate his returns (over Medallion's). We use a hypothetical example to illustrate the effect of fees. Assume a 50% stake for Simons, a 10 billion size for Medallion, an annual trading profit of 60%, and a 40% cut for the profit.[5] All investors, including early investors or employees not working for the firm anymore, will share 3.6 billion, which leaves Simons with 1.8 billion. Current employees will share 2.4 billion, which gives Simons 1.68 billion, assuming Simons has 70% equity among current Employees. As a result, Simons would earn 3.48 billion (not the 3 billion from the overall 50% stake), or a return of 69.6%. This implies that the net returns of Medallion might be lower than the returns of Simons. Therefore, it may probably be correct to conclude that the annualized compounded return of Medallion is less than 35%.

**Conclusions**

This paper shows that the naïve compounded return, obtained by compounding yearly net returns, turns out to be over 60% for the Medallion fund and is incorrect. This is because the value process of a fund, such as Medallion, is discontinuous.

A better estimation for the compounded return can be obtained by considering the fund sizes and trading profits. This leads to an estimated annualized compounded return of

[5] We ignore the 5% annual expense here.

31.8% for Medallion (before fees) for the 31 years up to 2018. Alternatively, the wealth growth of the manager can be utilized as a proxy for the compounded return of a fund. For Simons, the estimated annualized compounded growth is 25.6% for the 33 years up to 2020. Therefore, the annualized compounded return of Medallion is improbable to be more than 35% for the same period.

## References

Cornell, Bradford. 2020. Medallion Fund: The Ultimate Counterexample? The Journal of Portfolio Management. 46, 156-159.

Cochrane, John H. 2005. Asset Pricing, rev. ed. Princeton University Press, Princeton.

Guo, Shuxin, and Qiang Liu. 2024. Data-Generating Process and Time-Series Asset Pricing. arXiv:2405.10920.

Zuckerman, Gregory. 2019. The Man Who Solved the Market: How Jim Simons Launched the Quant Revolution. Portfolio, Penguin Group.

Table 1. Fund sizes and trading profits of Medallion from 1988 to 2018.

| Year No. | Fund Size | Trading Profit | Return (%) | Year No. | Fund Size | Trading Profit | Return (%) |
|---|---|---|---|---|---|---|---|
| 1 | 20 | 3 | 15.00 | 17 | 5,200 | 2,572 | 49.46 |
| 2 | 20 | 0 | 0.00 | 18 | 5,200 | 2,999 | 57.67 |
| 3 | 30 | 23 | 76.67 | 19 | 5,200 | 4,374 | 84.12 |
| 4 | 42 | 23 | 54.76 | 20 | 5,200 | 7,104 | 136.62 |
| 5 | 74 | 35 | 47.30 | 21 | 5,200 | 7,911 | 152.13 |
| 6 | 122 | 66 | 54.10 | 22 | 5,200 | 3,881 | 74.63 |
| 7 | 276 | 258 | 93.48 | 23 | 10,000 | 5,750 | 57.50 |
| 8 | 462 | 244 | 52.81 | 24 | 10,000 | 7,107 | 71.07 |
| 9 | 637 | 283 | 44.43 | 25 | 10,000 | 5,679 | 56.79 |
| 10 | 829 | 261 | 31.48 | 26 | 10,000 | 8,875 | 88.75 |
| 11 | 1,100 | 628 | 57.09 | 27 | 9,500 | 7,125 | 75.00 |
| 12 | 1,540 | 549 | 35.65 | 28 | 9,500 | 6,582 | 69.28 |
| 13 | 1,900 | 2,434 | 128.11 | 29 | 9,500 | 6,514 | 68.57 |
| 14 | 3,800 | 2,149 | 56.55 | 30 | 10,000 | 8,536 | 85.36 |
| 15 | 5,240 | 2,676 | 51.07 | 31 | 10,000 | 7,643 | 76.43 |
| 16 | 5,090 | 2,245 | 44.11 | | | | |

*Notes*: Fund size and trading profits before fees (in millions) are taken from Zuckerman (2019). Return: the ratio of trading profit to fund size. Year No. 1 is 1988, and Year No. 31 is 2018.

Table 2. Adjusted profits and year-end values from 1988 to 2018.

| Year No. | Adj. Profit | Ending Value | Gross Return | Year No. | Adj. Profit | Ending Value | Gross Return |
|---:|---:|---:|---:|---:|---:|---:|---:|
| 1 | 23 | 23 | 1.150 | 17 | 2,572 | 14,362 | 1.218 |
| 2 | 0 | 23 | 1.000 | 18 | 2,999 | 17,361 | 1.209 |
| 3 | 15 | 38 | 1.667 | 19 | 4,374 | 21,735 | 1.252 |
| 4 | 23 | 61 | 1.600 | 20 | 7,104 | 28,839 | 1.327 |
| 5 | 31 | 92 | 1.501 | 21 | 7,911 | 36,750 | 1.274 |
| 6 | 59 | 151 | 1.640 | 22 | 3,881 | 40,631 | 1.106 |
| 7 | 176 | 327 | 2.163 | 23 | 5,222 | 45,853 | 1.129 |
| 8 | 244 | 571 | 1.747 | 24 | 7,107 | 52,960 | 1.155 |
| 9 | 283 | 854 | 1.496 | 25 | 5,679 | 58,639 | 1.107 |
| 10 | 261 | 1,115 | 1.306 | 26 | 8,875 | 67,514 | 1.151 |
| 11 | 622 | 1,737 | 1.558 | 27 | 7,125 | 74,639 | 1.106 |
| 12 | 549 | 2,286 | 1.316 | 28 | 6,582 | 81,221 | 1.088 |
| 13 | 2,434 | 4,720 | 2.065 | 29 | 6,514 | 87,735 | 1.080 |
| 14 | 2,149 | 6,869 | 1.455 | 30 | 8,536 | 96,271 | 1.097 |
| 15 | 2,676 | 9,545 | 1.390 | 31 | 7,643 | 103,914 | 1.079 |
| 16 | 2,245 | 11,790 | 1.235 | | | | |

*Notes*: Adj. profits, which are adjusted according to Formula 2, and Ending values, which are the sums of the fund size of 20 in 1998 and all the profits up to the current year, (in millions). Gross return: the ratio of the current value to its previous value or 20 for the first year. Year No. 1 is 1988, and Year No. 31 is 2018.

Table 3. Annualized compounded returns of Medallion for various sub-periods.

| Period | 2002-2009 | 2010-2018 | 1988-2001 | 2002-2018 | 1988-2018 |
|---|---|---|---|---|---|
| Number of years | 8 | 9 | 14 | 17 | 31 |
| Initial size | 5,240 | 10,000 | 20 | 5,240 | 20 |
| Total trading profit | 33,762 | 63,283 | 6,869 | 97,045 | 103,914 |
| Final value | 39,002 | 73,283 | 6,889 | 102,285 | 103,934 |
| Compounded return (%) | 28.5 | 24.8 | 51.8 | 19.1 | 31.8 |

*Notes*: Value, size, and profit are in millions.